\documentclass[11pt]{article}
\usepackage{amsmath,amssymb,color,graphics,epsfig,cite}

\usepackage[unicode=false,
bookmarks=true,bookmarksnumbered=false,bookmarksopen=false,
breaklinks=false,pdfborder={0 0 1},backref=false,colorlinks=false]
{hyperref}
\usepackage{amsfonts}

\newcommand{\hoch}[1]{$\, ^{#1}$}

\newcommand{\be}{\begin{equation}}
\newcommand{\ee}{\end{equation}}
\newcommand{\bea}{\setlength\arraycolsep{2pt} \begin{eqnarray}}
\newcommand{\eea}{\end{eqnarray}}
\newcommand{\nn}{\nonumber}

\def\ft#1#2{{\textstyle{\frac{\scriptstyle #1}{\scriptstyle #2} } }}
\def\fft#1#2{{\frac{#1}{#2}}}

\def\0{{\sst{(0)}}}
\def\1{{\sst{(1)}}}
\def\2{{\sst{(2)}}}
\def\3{{\sst{(3)}}}
\def\4{{\sst{(4)}}}
\def\5{{\sst{(5)}}}
\def\6{{\sst{(6)}}}
\def\7{{\sst{(7)}}}
\def\8{{\sst{(8)}}}
\def\sst#1{{\scriptscriptstyle #1}}

\begin{document}

\begin{center}
{\Large {\bf Charging the Love numbers: Charged scalar response coefficients of Kerr-Newman black holes }}

\vspace{20pt}

Liang Ma\hoch{1}, Ze-Hua Wu\hoch{1}, Yi Pang\hoch{1,2} and  H. L\"{u}\hoch{1,2,3}

\vspace{5pt}

{\it \hoch{1}Center for Joint Quantum Studies and Department of Physics,\\
School of Science, Tianjin University, Tianjin 300350, China }

\vspace{5pt}

{\it \hoch{2}Peng Huanwu Center for Fundamental Theory, Hefei, Anhui 230026, China}

\vspace{5pt}

{\it \hoch{3}Joint School of National University of Singapore and Tianjin University,\\
International Campus of Tianjin University, Binhai New City, Fuzhou 350207, China}


\vspace{40pt}

\underline{ABSTRACT}
\end{center}

We study the electrically-charged massless scalar response to the general Kerr-Newman black hole and obtain the exact (complex) response coefficients, whose real parts are the Love numbers. We find there is a curious discontinuity between the neutral and charged scalar. The Love numbers vanish for the neutral scalar, but are inversely proportional to the electric charge of the scalar in the small charge limit. We study the near-zone of the dynamical equation with non-vanishing frequency $\omega$, and show that the system have an $SL(2,\mathbb{R})$ Love-like symmetry. The symmetry selects a special frequency $\omega_{\rm cr}$, which vanishes in the neutral case, so that the real part of the response coefficient vanishes identically. We also generalize the results into higher dimensions to some extend.

\vfill {\footnotesize {liangma\_physics\_0@tju.edu.cn\ \ \ wuzh\_2022@tju.edu.cn \ \ \ pangyi1@tju.edu.cn \ \ \ mrhonglu@gmail.com}}


\thispagestyle{empty}
\pagebreak


\tableofcontents
\addtocontents{toc}{\protect\setcounter{tocdepth}{2}}


\section{Introduction}

A solid body's response to a weak and slowly varying gravitational tidal force is captured by the tidal Love number. In the framework of general relativity (GR), the tidal effect is reflected in the analog of Newtonian potential $h_{00}=1+g_{00}$, which admits a mutipolar expansion of the form \cite{Binnington:2009bb,Landry:2015zfa,LeTiec:2020bos,Yunes:2005ve,Kol:2011vg}
\bea
h_{00} &=& -\sum_{\ell=2}\sum_{m=-\ell}^{\ell}\frac{(\ell-2)!}{\ell!}Y_{\ell m}\mathcal{E}_{\ell m}r^\ell\Big\{
\Big[1+\sum_{i=1}^{\infty}a_i\Big(\frac{r_s}{r}\Big)^i\Big]
+k_{\ell m}^{(0)}\Big(\frac{r}{r_s}\Big)^{-2\ell-1}
\Big[1+\sum_{i=1}^{\infty}b_i\Big(\frac{r_s}{r}\Big)^i\Big]
\Big\}\nn\\
&&+\hbox{complex conjugate.}\label{static external perturbation}
\eea
Here $\ell$ is the multipolar index, $m$ is the azimuthal harmonic number (magnetic number) obeying $\lvert m\rvert\leq\ell$ and ${\cal E}_{\ell m}$ denotes the tidal moment. Each coefficient $k_{\ell m}^{(0)}$ is the Love number, which when multiplied by the tidal moment, is proportional to multipole moment of the mass density. For a spinning body with rotation parameter $\Omega$ and low frequency $\omega\sim 0$, the Love number $k_{\ell m}^{(0)}$ is in general complex
\be
k_{\ell m}^{(0)}=\kappa_{\ell m}+i\nu_{\ell m}(\omega-m\Omega)+{\cal O}(\omega^2)\,,
\label{Kerr dissipation}
\ee
in which the real part $\kappa_{\ell m}$ describes the conservative tidal response while the imaginary part encodes the dissipation. For this reason, $\kappa_{\ell m}$ plays the role of the ordinary Love number and $\nu_{\ell m}$ is termed as the dissipation number (dissipation response coefficient) \cite{Chia:2020yla, Goldberger:2020fot} that captures the time lag between the external tidal field and the body's response due to dissipation. It has been shown that the tidal deformation can be revealed by the shape and phasing gravitational waves \cite{Flanagan:2007ix, Hinderer:2007mb}.

The expression in \eqref{static external perturbation} also demonstrates that in GR, both the source (tidal force) and the response come with an infinite series expansion in $1/r$. For a given $\ell$, the ``1" in the source term corresponds to the Newtonian limit and ($a_i$, $b_i$) are generated by nonlinear effects of GR. In fact, the infinite expansion of the source term can coalesce with the  subleading terms in the source term with the same power of $1/r$, leading to ambiguities in identifying the response coefficients alone.  One strategy avoiding this issue is to first analytically continue $\ell$ from integer to real number which prevents the hypergeometric function from reducing to finite polynomials \cite{Landry:2015zfa,Chia:2020yla,Charalambous:2021mea}. Some other proposals for static black holes can be found in \cite{Hui:2020xxx}.

We recall that for Ricci-flat Kerr black holes, to obtain the tidal Love numbers, one needs to solve the Teukolsky equations \cite{Teukolsky:1973ha} satisfied by components of the Weyl tensor. However, for more general  non-Ricci-flat rotating black holes such as Kerr-Newman black holes or black holes in supergravities, the generalizations of Teukolsky equations are either not available for the time being or become rather complicated. Nevertheless, one can still gain useful insights by studying lower spin analogs of Love numbers, namely, the response of vector or scalar field to external tidal field in the background of rotating black holes. Previous works \cite{Damour:2009vw, Binnington:2009bb, Fang:2005qq, Kol:2011vg, Chakrabarti:2013lua, Gurlebeck:2015xpa, Hui:2020xxx, Charalambous:2021mea, Cvetic:2021vxa} have confirmed that the spin-0 and spin-1 analogs of Love numbers exhibit similar features as the spin-2 tidal Love numbers. For instance, it is known that the real part of the spin-2 and lower spin tidal Love number vanishes for Kerr black holes in $D=4$. Of course, one should also be aware of some subtle differences between the spin-2 and lower spin tidal Love numbers. For instance, there exist magnetic-type responses in the spin-1 and spin-2 tidal Love numbers which are related to the odd-parity magnetic-type field strength (for spin-1) and Weyl tensor (for spin-2) \cite{Binnington:2009bb, Zhang:1986cpa, Charalambous:2024gpf} respectively. However, there is no magnetic-type response in the spin-0 case. Owing to the electromagnetic duality, the electric-type response is identical to the magnetic-type \cite{LeTiec:2020spy, LeTiec:2020bos, Charalambous:2021mea}. Beyond the linear response theory, nonlinearities in the tidal Love numbers of neutral black holes were considered in \cite{DeLuca:2023mio}.

In this work, we shall study the tidal response of a charged scalar field in the background of a Kerr-Newman black hole
\be
D_\mu D^\mu \Phi=0,\quad D_\mu=\partial_\mu-iq_e A_\mu\ .
\label{charged scalar}
\ee
This generalizes previous discussions \cite{Hui:2020xxx, Charalambous:2023jgq, Rodriguez:2023xjd, Cvetic:2021vxa} which only considered a neutral scalar field in  the background of a charged black hole.  Clearly, the charged scalar field can no longer mimic the behavior of the gravitational field; however, there are at least three advantages to consider a charged scalar. First of all, in the Newtonian potential, terms associated with the tidal source automatically disentangle with those associated with the response. The reason is that the source term begins with $r^L$, while the response term starts as $r^{-L-1}$ for
\be
L=\sqrt{(\ell +\ft12)^2- q_e^2 Q^2}-\fft12\,,\label{modifiedL}
\ee
which is a real number for generic $q_e$ and the black hole electric charge $Q$. Thus introducing a non-zero $q_e$ provides an alternative way of analytically continuing $\ell$ from integer to rational number. Secondly, the charged scalar field may mimic charged matter field around a Kerr-Newman black hole. The response coefficients computed here can thus capture, in the background of Kerr-Newman black holes, how the charged matter redistributes under the influence of other charged matter.  Finally, charged matter arises naturally in fundamental theories such as string theory as part of the supermultiplets and it is of theoretical interest to study their tidal force response. In this paper we use the charged massless scalar \eqref{charged scalar} as a toy model. One important result is that the charged Love number no longer vanishes, indicating that the neutrality may play an important role in its vanishing. In fact we find that for charged scalar, the vanishing of the charged spin-0 Love number occurs at some critical $\omega_{\rm cr}$, which happens to be zero for the neutral case.

The paper is organized as follows. In section 2, we focus on the discussion of the response function of a massless charged scalar in the background of Kerr-Newman black hole. In section 3, we study the $SL(2,\mathbb R)$ symmetry in the near-zone equation for the dynamical solution with non-zero frequency. We shall see that owing to this symmetry, the real part of the scalar response vanishes for the pseudo-static solutions. In section 4, we discuss charged scalar response in general dimensions, focusing on the Reissner-Nordstr\"om (RN) black hole backgrounds. We conclude the paper in section 5. In appendix, we show that in the extremal RN background, the charged massive scalar equation can be solved dynamically in terms of double confluent Heun's functions.

\section{Love number of charged scalar}

In this paper, we consider Einstein-Maxwell theory coupled to a massless charged scalar. The theory admits scalar hairless charged rotating black hole, namely the
Kerr-Newman black hole
\bea
ds^2&=&-\frac{\Delta-a^2\sin^2\theta}{\rho^2}\left[dt+\frac{(2\mu r-Q^2)a\sin^2\theta}{\Delta-a^2\sin^2\theta}d\varphi
\right]^2+\rho^2\left[\frac{dr^2}{\Delta}+d\theta^2+
\frac{\Delta\sin^2\theta}{\Delta-a^2\sin^2\theta}d\varphi^2
\right],\cr
A_{(1)}&=&-\frac{Qr}{\rho^2}(dt-a\sin^2\theta d\varphi)\,,\qquad
\Delta=r^2-2\mu r+a^2+Q^2\,,\qquad \rho^2=r^2+a^2\cos^2\theta\,.
\eea
The solution contains three integration constants, parameterizing the mass $M=\mu$, angular momentum $J=Ma$ and the electric charge $Q$. For sufficiently large mass, the solution described a black hole with both inner $r_-$ and outer $r_+$ horizons
\be
r_\pm=\mu\pm\sqrt{\mu^2-a^2-Q^2}\,.
\ee
The remaining thermodynamic quantities associated with both inner and outer horizons are
\be
\Omega_\pm=\frac{a}{r_\pm^2+a^2}\,,\qquad \Phi_e^\pm=\frac{Qr_\pm}{r_\pm^2+a^2}\,,\qquad T_\pm=\fft{\kappa_\pm}{2\pi}=\frac{r_\pm-r_\mp}{4\pi(r_\pm^2+a^2)}\,,\qquad
S_\pm =\pi (r_\pm^2 + a^2)\ ,
\ee
which satisfy the first laws $dM=T_\pm dS_\pm + \Omega_\pm dJ + \Phi_e^\pm dQ$, associated with both outer and inner horizons, respectively.

The static ($\omega=0$) scalar field equation \eqref{charged scalar} can be solved by separation of variables in which the angular part of the scalar field is expanded in terms of spin-0 spherical harmonics
\be
\Phi=\sum_{\ell, m}\mathcal{E}_{\ell m}^{(0)}\,R_{\ell m}(r)\,Y_{\ell m}(\theta,\varphi)\,.
\ee
The radial function $R_{\ell m}$ satisfies
\be
\partial_r(\Delta\partial_r R_{\ell m})- \left[\ell  (\ell +1)-\frac{\left( ma+q_eQ r \right){}^2}{\Delta }\right]R=0\,.
\ee
It is advantageous to introduce a new radial variable
\be
x\equiv\frac{r-r_+}{r_+-r_-}\,,\qquad \rightarrow \qquad x\in [0,\infty)\,,\label{coord x}
\ee
in terms of which, the radial equation takes the form
\be
x (x+1) R_{\ell m}''(x)+(2 x+1) R_{\ell m}'(x)- \left[\ell (\ell+1)-\frac{m^2 (\alpha +\beta  x)^2}{x (x+1)}\right]R_{\ell m}(x)=0\,.\label{radial equations x}
\ee
We see that all the black hole and scalar information is encoded in two parameters $(\alpha, \beta)$, given by
\be
\beta = \frac{ q_eQ}{m}\,,\qquad
\alpha = \frac{\beta r_++ \sqrt{r_+r_- -Q^2}}{r_+-r_-}= \fft{1}{4\pi m T_+} (q_e \Phi_e^+ + m \Omega_+)\,.\label{Kerr-Newman alphabeta}
\ee
The case with $q_e=0$ reduces to the previously known examples in literature. For nonzero $q_e$, a new term proportional to $\beta$ is appears in the square bracket in \eqref{radial equations x}. In particular, in the asymptotic $x\rightarrow \infty$ region, the term inside the square bracket is no longer simply $\ell(\ell+1)$, but it is modified to
\be
\ell(\ell+1) - \beta^2 m^2= L(L+1)\,,
\ee
where the modified ``multipolar index $L$'' is given by \eqref{modifiedL}. Thus the parameter $\beta$ alters the equation at asymptotic region by modifying the multipolar index. Needless to say, it changes the equation at the intermediate and horizon regions as well.

It is rather remarkable that the equation \eqref{radial equations x} can still be solved analytically in terms of hypergeometric functions. Specifically, it is a linear combination of two independent branches
\bea
R_{\ell m}(x) &=& c_1 R_{\ell m,1}(x)+c_2 R_{\ell m,2}(x)\,,\label{static solution 1}\\
R_{\ell m,1}(x)&=&\fft{x^{i  m\alpha }}{(1+x)^{i m (\alpha -\beta )}} \, _2F_1\left(a_e,b_e;1+2 i  m\alpha;-x\right)\,,\nn\\
R_{\ell m,2}(x)&=&\fft{x^{-i m \alpha }}{(1+x)^{i m (\alpha -\beta )}} \, _2F_1\left(a_e-2 i m \alpha ,b_e-2 i m \alpha ;1-2 i  m\alpha;-x\right)\,,
\eea
where the two parameters are given by
\be
a_e= i \beta m -L\,,\qquad
b_e= i \beta m + L+1\,,\qquad c_e=1+2 i  m\alpha\,.\label{static solution 2}
\ee
For the neutral external perturbation $q_e=0$, $R_{\ell m,1}$ reduces to the Kerr black hole case, where the second branch should be discarded by setting $c_2=0$ \cite{Charalambous:2021mea}. There are two reasons behind this choice. One is that on horizon $x=0$, after removing the frame-dragging factor $(x/(1+x))^{i  m\alpha }$ \cite{Charalambous:2021mea}, the $R_{\ell m,1}$ is regular, but $R_{\ell m,2}$ has a branch-cut singularity since the $x$-derivative of the quantity,
\be
x^{-i  m\alpha}R_{\ell m,2}\sim x^{-2i  m\alpha}\,,
\ee
is singular \cite{LeTiec:2020bos}. Alternatively, we can consider the more general time-dependent solutions with frequency $\omega$. The general solution can be decomposed into the ingoing and outgoing branches. The boundary condition on the black hole background selects the ingoing condition. Having done this, one can then set $\omega=0$ and obtain the static solution with appropriate boundary condition. It turns out that for $q_e=0$, this process selects $R_{\ell m,1}$ \cite{LeTiec:2020bos}. For our more general $q_e\ne 0$ solution, we see that only the parameter $\alpha$ enters the leading order terms in the horizon expansion. It follows that the proper boundary condition selects $R_{\ell m,1}$ as well. We therefore set $c_2=0$ from now on.

To extract the scalar response coefficients, we need to expand the radial solution at the asymptotic infinity. Utilizing the analytic continuation of hypergeometric functions at $x=\infty$
\bea
{}_2F_1(a,b;c;x)&=&\frac{\Gamma (c) \Gamma (b-a)}{\Gamma (b) \Gamma (c-a)}(-x)^{-a} \, _2F_1\left(a,a-c+1;a-b+1;x^{-1}\right)\cr
&&+\frac{\Gamma (c) \Gamma (a-b)}{\Gamma (a) \Gamma (c-b)}(-x)^{-b} \, {}_2F_1\left(b,b-c+1;b-a+1;x^{-1}\right),
\eea
we find the original multipolar index $\ell$ is now replaced by $L$, with the asymptotic behavior
\be
R_{\ell m,1}(x)\sim\#_1\, x^{L}+\#_2\,x^{-L-1}\,.
\label{boundary condition KN}
\ee
We are interested in the case of small scalar charge $q_e$, i.e.~$
(2 \ell+1)^2- 4 q_e^2 Q^2 \geq0$, so that the modified multipolar index $L$ is always real. Unlike the neutral case, where $L=\ell>0$, the charged solution can have two distinct sectors: (1) $L>0$ and (2) $-\fft12 <L<0$. We discuss them separately.

\noindent {\bf case 1:} $L>0$. In this case,  the term $\#_1\,x^{L}$ in \eqref{boundary condition KN} is the growing mode. We can set
\be
c_1=\frac{\Gamma (1+L+i m \beta ) \Gamma (1+L+i m (2 \alpha -\beta ))}{\Gamma (2 L+1) \Gamma (1+2 i m \alpha )}\,,
\ee
to fix the normalization at $x=\infty$. We then have $R_{\ell m,1}(x) \sim x^{L}+k_{\ell m}^{(0)}x^{-L-1}$, where
\be
k_{\ell m}^{(0)}=\frac{\Gamma (-2 L-1) \Gamma (1+L+i m \beta ) \Gamma (1+L+i m (2 \alpha -\beta ))}{\Gamma (2 L+1) \Gamma (-L+i m \beta ) \Gamma (-L+i m (2 \alpha -\beta ))}\,,\qquad L>0\,.\label{k for KN}
\ee
This expression\footnote{Following the expression of gravitation potential at large distance \eqref{static external perturbation}, the definition of response coefficient should be in the coordinate $r$. Here we drop the overall factor $\left(\frac{r_+-r_-}{r_s}\right)^{2L+1}$ for convenient.} illustrates how the scalar charge $q_e$ modifies the argument of each $\Gamma$ function, by changing $\ell$ to $L$, as well as modifying the imaginary part of these quantities. One important effect is that $L$ is no longer an integer, even if the scalar charge $q_e$ obeys certain quantization rule. Hence, $k_{\ell m}^{(0)}$ manifests as a complex number with both real and imaginary parts. We can make use of the mirror formula $\Gamma(z)\Gamma(1-z)=\frac{\pi}{\sin\pi z}$ to write real and imaginary parts
\be
k_{\ell m}^{(0)}=\kappa_{\ell m}^{(0)}+i\nu_{\ell m}^{(0)}\,.
\ee
To present the explicit results, it is advantageous to introduce an overall real quantity
\be
A_{\alpha,\beta}=\frac{\Gamma (-2 L-1)  }{2 \pi ^2 \Gamma (2 L+1)}\vert \Gamma (1+L+i m \beta )\vert^2\,  \vert\Gamma (1+L+i m (2 \alpha -\beta )) \vert^2\,.
\ee
The Love number $\kappa_{\ell m}^{(0)}$ and the dissipation response coefficient $\nu_{\ell m}^{(0)}$ are then given by
\bea
\kappa_{\ell m}^{(0)}&=&\Big[\cosh (2 \pi  m (\alpha-\beta ))-\cos (2 \pi  L) \cosh (2 \pi m \alpha  )\Big]A_{\alpha,\beta}\,,\nn\\
\nu_{\ell m}^{(0)}&=&-\sin (2 \pi  L) \sinh (2 \pi m \alpha  )A_{\alpha,\beta}\,.\label{genres}
\eea

Having obtained the general results for the charged scalar, we now study various limits. Setting $q_e=0$ leads to $\beta=0$ and $L=\ell$, which leads to the previously known result, namely \cite{Chia:2020yla, Goldberger:2020fot, Poisson:2020mdi, Poisson:2020vap, Charalambous:2021mea}
\be
\kappa_{\ell m}^{(0)}=0\,,\qquad \nu_{\ell m}^{(0)}=-\frac{(\ell!)^2m\alpha}{(2\ell+1)!\,(2\ell)!}\prod_{n=1}^{\ell}
\left(n^2+4m^2\alpha^2\right).\label{qe=0case}
\ee
In other words, the Love number vanishes, with a nonzero dissipation coefficient. In obtaining this result, a subtle regularization was used, by taking $\ell\rightarrow \ell + \varepsilon$ and then sending $\varepsilon\rightarrow 0$. This has the effect of regularize the divergence of $\Gamma(-\ell)$ for integer $\ell$ as follows
\be
\Gamma(-\ell)\rightarrow\Gamma(-\ell-\varepsilon)=
\frac{(-1)^{\ell+1}}{\ell!\,\varepsilon}\,.
\ee
One needs to further make use of the identity
\be
\lvert \Gamma(1+\ell+iB) \rvert^2=\frac{\pi B}{\sinh(\pi B)}\prod_{n=1}^{\ell}\left(n^2+B^2\right).
\ee
We now consider small $q_e$, in which case, we have
\be
L= \ell + \delta\ell \,,\qquad \delta\ell= - \fft{m^2 \beta^2}{2\ell+1}\,.
\ee
The leading term of $A_{\alpha,\beta}$ is actually divergent, given by
\be
A_{\alpha,\beta}\sim \fft{1}{\delta\ell}\, \frac{(\ell!)^2}{2\pi (2\ell)!(2\ell+1)!} \frac{ m\alpha}{\sinh(2\pi m\alpha)}
\prod_{n=1}^{\ell}\left(n^2+4m^2\alpha^2\right) + \cdots\,.
\ee
The coefficients in \eqref{genres} in the small $q_e$ or $\beta$ expansion are
\bea
\cosh (2 \pi  m (\beta -\alpha )) - \cos (2 \pi  L) \cosh (2 \pi m \alpha  ) &\sim &
-2   \pi  m\beta \sinh (2 \pi m\alpha)\,,\nn\\
-\sin (2 \pi  L) \sinh (2 \pi m \alpha  ) &\sim&
-2 \pi\delta\ell\sinh (2 \pi m \alpha  )\,.
\eea
This implies that the leading-order of $\nu_{\ell m}^{(0)}$, the imaginary part, is unchanged, given by \eqref{qe=0case}. The real part, on the other hand, has the leading-order dependence
\be
\kappa_{\ell m}^{(0)}=\frac{(1+2 \ell ) (\ell !)^2}{(2 \ell )! (1+2 \ell )!}\frac{\alpha }{\beta }\prod_{n=1}^{\ell}\left(n^2+4m^2\alpha^2\right) + \cdots\,.\label{kappadiv}
\ee
Thus we see that there is a discontinuity of the Love number about the scalar charge $q_e=0$. It vanishes for $q_e=0$, but it is inversely proportional to $q_e$ for small but non-vanishing  $q_e$, and hence it diverges in the $q_e\rightarrow 0$ limit. Mathematically, it may related to the fact that introducing the scalar charge alters the asymptotic behavior of the equation \eqref{radial equations x}; however, it is difficult to understand this physically. One resolution of this discontinuity may appeal to the charge quantization of $q_e$, which puts an upper bound on the charged Love number. A similar discontinuity was also found in the massive perturbations on the BTZ black hole background case \cite{DeLuca:2024ufn}  where the mass plays the role of $q_e$.

In the above calculation, we have assumed that $\ell$ is simply an integer. As we have seen earlier in the $q_e=0$ case, it is advantageous to analytically extend $\ell$ to a real number by $\ell\rightarrow \ell + \varepsilon$ and then set $\varepsilon=0$ as an appropriate limit. We can do the same thing for non-vanishing $q_e$, and we find
\be
\kappa_{\ell m}^{(0)}\sim\frac{\pi   \cosh (2 \pi    m\alpha)\varepsilon ^2-  m \sinh (2 \pi   m \alpha  )\beta}{\varepsilon -\frac{ m^2}{2 \ell +1}\beta ^2}\frac{m\alpha  }{\sinh (2 \pi  m\alpha  )}\frac{(\ell !)^2}{(2 \ell +1)! (2 \ell )!}\prod_{n=1}^{\ell}\left(n^2+4m^2\alpha^2\right).
\ee
At first sight, the final answer depends on how the two parameters $\varepsilon$ and $\beta$ approach zeros. However, $\varepsilon$ is a regularization parameter that must be set to zero whilst $q_e$ (or $\beta$) is a physical charge parameter that can be small but set to a non-vanishing value. Thus the result \eqref{kappadiv} is independent of the regularization scheme.

Another special case worth mentioning is that when we take $L\in\mathbb{N}$. The corresponding $k_{\ell m}^{(0)}$ would lead to an infinity
\be
\frac{k_{\ell m}^{(0)}}{\Gamma (-2 L-1)}=-\frac{m^2(2\alpha-\beta)\beta}{(2L)!}
\prod_{n_1=1}^L[n_1^2+m^2(2\alpha-\beta)^2]\prod_{n_2=1}^L[n_2^2+m^2\beta^2]\,,\qquad L>0\,.
\ee
This is analogous to the running Love number case discussed in \cite{Kol:2011vg}. When $L$ is an integer, the descendants of the growing mode would coalesce the decaying responsive mode; consequently logarithmic terms emerge. We find that the appropriate Love number becomes logarithmic divergent:
\be
k_{\ell m}^{(0)}=-\frac{m^2(2\alpha-\beta)\beta}{2(2L)!(2L+1)!}
\prod_{n_1=1}^L[n_1^2+m^2(2\alpha-\beta)^2]
\prod_{n_2=1}^L[n_2^2+m^2\beta^2]\log\frac{r_0}{r}\,,\qquad L>0\,.
\ee
where the length scale $r_0$ is a renormalization scale to be fixed by experiments, which we expect to be of $\mathcal{O}(r_s)$.

\noindent{\bf Case 2}: $-\frac{1}{2}<L<0$. In this case, the solution $R_{\ell m,1}(x)$ does not have a growth mode. We can thus have an alternative normalization condition
\be
c_1=\frac{\Gamma (-L+i m \beta ) \Gamma (-L+i m (2 \alpha -\beta ))}{\Gamma (-2 L-1) \Gamma (1+2 i m \alpha )}\,,
\ee
so that we have $R_{\ell m,1}(x)\sim x^{-L-1}+k_{\ell m}^{'(0)}x^{L}$ as well. Thus the response coefficient can be either $k_{\ell m}^{(0)}$, formally given by
\eqref{k for KN} but with $-\fft12<L<0$, or its inverse.

\section{Love Symmetry}

The fact that Love number vanishes for 4D black hole suggests the presence of some hidden symmetry. In \cite{Charalambous:2021kcz,Charalambous:2022rre}, ``Love symmetry'' was proposed to explain this mysterious vanishing of the Love numbers. Further understandings of the origin of this symmetry can be found in \cite{BenAchour:2022uqo, Sharma:2024hlz}.
To understand this symmetry, one needs to put the $\omega=0$ solution in the larger solution space with $\omega\ne 0$. In the appropriate near-zone approximation, there exists an $SL(2, \mathbb R)$ symmetry, dubbed as the Love symmetry, that governs the black hole perturbations. It was shown in \cite{Charalambous:2021kcz, Charalambous:2022rre}, if the static ($\omega=0$) perturbations belong to the highest weight representation of Love symmetry, then the Love number vanishes.

In our charged scalar perturbation \eqref{charged scalar}, solutions with non-vanishing $\omega$ take the form
\be
\Phi(t,r,\theta,\varphi)=\phi(t,r,\varphi)S(\theta)=e^{-i\omega t+im\varphi}R(r)S(\theta)\,.\label{charged dynamic scalar}
\ee
We choose the near-zone splitting similar to \cite{Charalambous:2021kcz, Charalambous:2022rre}
\bea
&&\partial_r(\Delta\partial_r R)+(V_0+\epsilon V_1)R=\ell(\ell+1)R\,,\nn\\
&&\frac{1}{\sin\theta}\partial_\theta(\sin\theta\partial_\theta S)+\left(\epsilon\omega^2a^2\cos^2\theta-\frac{m^2}{\sin^2\theta}
\right)S=-\ell(\ell+1)S\,,\nn\\
V_0&=&\frac{\left(r_+^2+a^2\right){}^2}{\Delta }\left[\left(\omega -m \Omega _H\right){}^2-4 m\omega  \Omega _H\frac{r-r_+ }{r_+-r_-}
+\frac{r q_e \Phi _e }{r_+^2}\left(r q_e \Phi _e+2 m  \Omega _Hr_+\right)
\right],\nn\\
V_1&=&\frac{2 \omega  \left(\mu  \left(ma \beta_\kappa  +4 \mu ^2 r_+ \omega \right)-Q^2 \left(ma +2 \mu  r_+ \omega \right)\right)}{\left(r-r_-\right) r_+}
+\omega ^2 \left(r^2+2 \mu  r+4 \mu ^2-Q^2\right)\cr
&&-\frac{2\omega q_e Q r   \left(r^2+a^2\right) }{\Delta }\,.\label{Teukolsky}
\eea
Here $\Omega_H=\Omega_+$, $\Phi_e=\Phi_e^+$ and $\beta_\kappa = 1/\kappa_+$. For non-zero $\omega$, the general solution cannot be solved. However, in the near-zone approximation $\epsilon=0$, the leading terms of the radial equation \eqref{Teukolsky} can again be solved in terms of hypergeometric function.

Here, we will discuss the near-zone approximation. In order to neglect the $V_1$ part in \eqref{Teukolsky}, we should impose the small frequency condition $\omega\mu,\ \omega a,\ \omega Q\ll1$ and the near region condition $\omega r\ll1$. Notice that the last term of $V_1$ diverges as $r\rightarrow r_+$. To avoid this divergence, the radial coordinate $r$ must be off the horizon by certain amount, specifically we find
\be
\omega r_+ q_eQ\ll \frac{r-r_+}{r_+},\quad r\sim r_+\,.
\ee
Further more, we also require the last term in $V_1$ to be negligible compared to the charged dependent terms in $V_0$, which implies
\be
|q_e\Phi_e\frac{r}{r_+}+2m\Omega_H|\gg2\omega\frac{r^2+a^2}{r_+^2+a^2}\,,
\ee
from which we derive the additional condition for the near-zone approximation
\be
q_eQ\gg\omega r\,.
\ee

After choosing the coordinate $x$ of \eqref{coord x}, the leading-term equation becomes
\be
x (x+1) R''(x)+(2 x+1) R'(x)+ \left(\frac{\mathcal{A}^2}{x}-\frac{\mathcal{B}^2}{x+1}\right)R(x)
=L(L+1)R(x)\,,\label{Teukolsky 1}
\ee
where $L$ is given by \eqref{modifiedL}, and it acts as the effective multipolar index as in \eqref{boundary condition KN}. The coefficients $({\cal A},{\cal B})$ are
\bea
\mathcal{A}&=&\frac{r_+^2+a^2}{r_+-r_-}\sqrt{\left(\omega -m \Omega _H\right){}^2+q_e \Phi _e \left(q_e \Phi _e+2 m \Omega _H\right)}\,,\cr
\mathcal{B}&=&\frac{r_+^2+a^2}{r_+ \left(r_+-r_-\right)}\sqrt{r_+^2 \left(\omega+m \Omega _H \right){}^2+ q_e \Phi _e r_-\left( q_e \Phi _er_-+2 m  \Omega _Hr_+\right)}\,.
\eea
The general solutions contain two linearly-independent branches
$R(x)=c_1 R_{1}(x)+c_2 R_{2}(x)$, with
\bea
R_{1}(x)&=&x^{i \mathcal{A}} (1+x)^{i \mathcal{B}} \, _2F_1(-L+i (\mathcal{A}+\mathcal{B}),1+L+i (\mathcal{A}+\mathcal{B});1+2 i \mathcal{A};-x)\,,\cr
R_{2}(x)&=&x^{-i \mathcal{A}} (1+x)^{i \mathcal{B}} \, _2F_1(-L-i (\mathcal{A}-\mathcal{B}),1+L-i (\mathcal{A}-\mathcal{B});1-2 i \mathcal{A};-x)\,.
\eea
It should be emphasized when $\omega=0$, the $\epsilon$ terms in \eqref{Teukolsky} all vanish. Thus although the near-zone solution is an approximate solution for any $\omega\ne 0$, it is the exact solution for $\omega=0$. Consequently, the above near-zone discussion is consistent with the exact static perturbation case. We can set $\omega=0$ to reproduce the static results (\ref{static solution 1},\ref{static solution 2}).

The general dynamical scalar response is
\be
k_{\ell m}^{(0)}(\omega)=\frac{\Gamma (-2 L-1) \Gamma (1+L+i (\mathcal{A}-\mathcal{B})) \Gamma (1+L+i (\mathcal{A}+\mathcal{B}))}{\Gamma (2 L+1) \Gamma (-L+i (\mathcal{A}-\mathcal{B})) \Gamma (-L+i (\mathcal{A}+\mathcal{B}))}\,.\label{electric Love number}
\ee
The fact that the near-zone equation can be solved by hypergeometric functions indicates that it has an $SL(2, \mathbb{R})$ symmetry, which we shall call it ``dynamical Love symmetry''. This symmetry allows the charged scalar Klein-Gordon equation to be rewritten as the Casimir eigenequation
\be
\mathcal{C}_2^{(L)}\phi^{(L)}(t,r,\varphi)=L(L+1)\phi^{(L)}(t,r,\varphi)\,,
\label{Teukolsky C}
\ee
where $\mathcal{C}_2^{(L)}$ and $\phi^{(L)}(t,r,\varphi)$ are the effective Casimir operator and effective wavefunction respectively.
The superscript $(L)$ denotes that the corresponding eigenvalue is $L(L+1)$.
Specifically, the $SL(2, \mathbb{R})$ generators take the form
\bea
\mathcal{C}_2^{(L)}&=&H_0^{(L)}H_0^{(L)}-\frac{1}{2}
\left(H_{+1}^{(L)}H_{-1}^{(L)}+H_{-1}^{(L)}H_{+1}^{(L)}
\right),\cr
H_0^{(L)}&=&-\beta_\kappa\partial_t\,,\qquad H_{\pm1}^{(L)}=e^{\pm\frac{t}{\beta_\kappa}}\left[
\mp\sqrt{\Delta}\partial_r+\partial_r(\sqrt{\Delta})
\beta_\kappa\partial_t+\frac{a}{\sqrt{\Delta}}\partial_\varphi
\right],
\label{Casimir and H}
\eea
which indeed fulfill the $SL(2, \mathbb{R})$ algebra
\bea
&&[H_m^{(L)},H_n^{(L)}]=(m-n)H_{m+n}^{(L)}\,,\qquad m,n=-1,0,+1\,,
\cr
&&[\mathcal{C}_2^{(L)},H_m^{(L)}]=0\ .
\eea

In \eqref{Teukolsky C}, for the radial part of $\phi^{(L)}$ to satisfy the same equation as that of $\phi$ \eqref{Teukolsky 1}, $\phi^{(L)}$ must acquire different frequency and azimuthal quantum number
\bea
&&\phi^{(L)}(t,r,\varphi)=e^{-i\omega^{(L)} t+im^{(L)}\varphi}R(r)\,,
\nn\\
\omega^{(L)}&=&\frac{r_+-r_-}{2(r_+^2+a^2)}(\mathcal{B}-\mathcal{A})\,,\qquad m^{(L)}=\frac{r_+-r_-}{2a}(\mathcal{A}+\mathcal{B})\,.
\label{electric Love generator}
\eea
It is important to note that the ``pseudo-static" $\phi^{(L)}$ with $\omega^{(L)}=0$ corresponds to a stationary $\phi$ with a non-vanishing frequency $\omega=\omega_{\rm cr}$, given by
\be
\omega^{(L)}=0\qquad \Rightarrow\qquad\omega_{\rm cr}=\frac{\left(r_+-r_-\right) q_e \Phi _e }{4 m  \Omega _Hr_+^2}\left[\left(r_++r_-\right) q_e \Phi _e+2 m  \Omega _Hr_+\right]\,.
\label{"static"}
\ee
The critical value $\omega_{\rm cr}$ vanishes when $q_e=0$, reproducing the static solution of $\phi$ discussed before.

As $L$ is generally not an integer, but a real number, the corresponding highest weight non-unitary representation is infinite dimensional. However, for certain choice of parameters, $L$ can be integer valued. In this case, the $SL(2, \mathbb{R})$ algebra \eqref{electric Love generator} can have finite dimensional non-unitary representations. For instance, the highest weight vector $v^{(L)}_{-L,0}$ has an imaginary frequency $\omega^{(L)}_{-L,0}=\frac{iL}{\beta_{\kappa}}$, and satisfies
\bea
&&H_{+1}^{(L)}v^{(L)}_{-L,0}=0\,,\qquad H_{0}^{(L)}v^{(L)}_{-L,0}=-Lv^{(L)}_{-L,0}\,,\cr
\Rightarrow&&
v^{(L)}_{-L,0}=\left(\frac{r-r_+}{r-r_-}\right)^{\frac{im^{(L)}a}{r_+-r_-}}
e^{im^{(L)}\varphi}\left(-e^{t/\beta_\kappa}\Delta^{1/2}\right)^L.
\label{highest weight}
\eea
Consequently, all the descendants generated by $H_{-1}^{(L)}$,
\be
v^{(L)}_{-L,n}=(H_{-1}^{(L)})^nv^{(L)}_{-L,0}\,,\qquad H_{0}^{(L)}v^{(L)}_{-L,n}=(n-L)v^{(L)}_{-L,n}\,,
\ee
are also solutions of the near-zone equations of motion (\ref{Teukolsky},\ref{Teukolsky 1},\ref{Teukolsky C}). The expression of $H_{0}^{(L)}$ \eqref{Casimir and H} implies that its eigenvalue is related to the frequency via
\be
H_{0}^{(L)}v^{(L)}_{-L,n}=i\beta_\kappa\omega^{(L)}_{-L,n}v^{(L)}_{-L,n}\,,\qquad \Rightarrow\qquad \omega^{(L)}_{-L,n}=i\frac{L-n}{\beta_\kappa}\,.
\ee
In particula, the pseudo-static solution $\Phi(\omega^{(L)}=0)$ is proportional to the lowest weight vector $v^{(L)}_{-L,L}=(H_{-1}^{(L)})^L\, v^{(L)}_{-L,0}$. Below we show that $v^{(L)}_{-L,L}$ does not have terms with the decaying power $r^{-L-1}$.  We first rewrite $v^{(L)}_{-L,L}$ as
\be
v^{(L)}_{-L,L}=\left(\frac{r-r_+}{r-r_-}\right)^{\frac{im^{(L)}a}{r_+-r_-}}
e^{im^{(L)}\varphi}F(r)\,.
\ee
From the expression of $H_{+1}^{(L)}$ given in \eqref{Casimir and H}, we observe that
\be
(H_{+1}^{(L)})^n\bigg[\left(\frac{r-r_+}{r-r_-}\right)^{\frac{im^{(L)}a}{r_+-r_-}}
e^{im^{(L)}\varphi}F(r)\bigg]
=\left(\frac{r-r_+}{r-r_-}\right)^{\frac{im^{(L)}a}{r_+-r_-}}
e^{im^{(L)}\varphi}
\Big(-e^{t/\beta_\kappa}\Delta^{1/2}\Big)^n\frac{d^n}{dr^n}F(r).
\ee
As the lowest weight vector is annihilated by $(H_{+1}^{(L)})^{L+1}$, we obtain
\be
(H_{+1}^{(L)})^{L+1}v^{(L)}_{-L,L}=
\left(\frac{r-r_+}{r-r_-}\right)^{\frac{im^{(L)}a}{r_+-r_-}}
e^{im^{(L)}\varphi}\left(-e^{t/\beta_\kappa}\Delta^{1/2}\right)^{L+1}
\frac{d^{L+1}}{dr^{L+1}}F(r)\,,
\ee
which implies the $L$-order polynomial $F(r)$. This concludes our proof that the pseudo-static solution with $L\in\mathbb{N}$ does not have terms with decaying powers $r^{-L-1}$. This $SL(2,\mathbb R)$ Love symmetry thus provides an explanation for the disappearance of dynamical Love numbers for the ``pseudo-static" case with $L\in\mathbb{N}$. The dynamical response coefficient \eqref{electric Love number} then reduces to pure dissipation response coefficient
\be
k_{\ell m}^{(0)}(\omega^{(L)}=0,L\in\mathbb{N})=-i\frac{\mathcal{A}+\mathcal{B}}{2}
\frac{(L!)^2}{(2L)!(2L+1)!}\prod_{n=1}^{L}\left[
n^2+(\mathcal{A}+\mathcal{B})^2
\right].
\ee
The derivation is analogous to the static case \cite{Charalambous:2021kcz, Charalambous:2022rre}, corresponding to $\omega_{\rm cr}=0$. Since the lowest-weight solution happens to be the pseudo-static solution $\Phi(\omega^{(L)}=0)\propto v^{(L)}_{-L,L}$, rather than a true static one $\Phi(\omega=0)$, we see that, for charged scalar fields, we cannot use Love symmetry to describe the static scalar response of charged black holes. This is consistent with the fact that the Love number does not vanish in this case. (However, it should be remarked here that for the the pseudo-static case with $\omega_{\rm cr}\ne 0$, the solution is valid only for the near-zone.)

Finally we remark that for half-integer $2L\in2\mathbb{N}+1$, the divergent behavior of \eqref{electric Love number} can be associated with RG running
\be
k_{\ell m}^{(0)}(\omega^{(L)}=0,2L\in2\mathbb{N}+1)
=\frac{(-1)^{2L}}{(2L)!(2L+1)!}\frac{\Gamma (1+L) \Gamma (1+L+i (\mathcal{A}+\mathcal{B}))}{ \Gamma (-L) \Gamma (-L+i (\mathcal{A}+\mathcal{B}))}\log\frac{r_0}{r}\,,
\ee
analogous to those discussed in \cite{Kol:2011vg}.

\section{$D$-dimensional RN black holes}

In previous sections, we focus our discussion on four dimensions. It is also of interest to generalize the results to higher dimensions. However, exact solutions of charged rotating black holes do not exist in Einstein-Maxwell theory in higher dimensions, but only in some supergravities whose structures depend on specific dimensions. In this section, we shall only consider the general $D$-dimensional (static) RN black holes:
\be
ds^{2} = - f(r) dt^{2} + \fft{1}{f(r)} dr^{2} + r^2 d \Omega_{D-2}^{2}\,,\qquad
A_{(1)} = \psi dt\,.
\ee
where the metric is given by
\bea
f(r) &=& 1-\fft{2 \mu}{r^{D-3}}+\fft{Q^{2}}{r^{2(D-3)}}=\left(1-\fft{\tilde{r}_{+}}{\tilde{r}}\right)
\left(1-\fft{\tilde{r}_{-}}{\tilde{r}}\right), \nn\\
\psi &=& - \sqrt{\fft{D-2}{2(D-3)}} \fft{Q}{r^{D-3}}\,,\nn\\
\tilde{r}_{\pm} &=& r_{\pm}^{D-3} = \mu \pm \sqrt{\mu^{2} - Q^{2}} \,,\qquad \tilde{r} = r^{D-3}\,.
\eea
The radial equation of the electric scalar field is
\bea
& &\Delta(r) \fft{d}{dr} \left( \Delta(r) \fft{d}{dr}R(r) \right) + U(r) R(r) = 0\,,\qquad
\Delta(r) = r^{D-2}f(r)\,,\nn\\
& &U(r) =  r^{2 (D-2)} (\omega + q_{e} \psi)^{2} - r^{2(D-3)} f(r) \left(\ell^{2} + (D-3)\ell + \mu_0^2 r^2\right).\label{eqRNfull}
\eea
For the massless $(\mu_0=0$) and static $(\omega=0)$ case, we redefine the radial coordinate by $x=\ft{\tilde{r}-\tilde{r}_{+}}{\tilde{r}_{+} - \tilde{r}_{-}}$, and the equation becomes
\be
x (x+1) R_{\ell m}''(x) + (2x+1) R_{\ell m}'(x)  - \left(\tilde{\ell}(\tilde{\ell}+1) - \fft{(\alpha + \beta x)^{\ft{2}{D-3}}}{x(1+x)} \right) R_{\ell m}(x) = 0\,.\label{eqRNq}
\ee
Where $\tilde{\ell} = \fft{\ell}{D-3}$ and
\be
\alpha = \fft{(q_{e} Q)^{D-3} \tilde{r}_{+}}{(\tilde{r}_{+} - \tilde{r}_{-})^{D-3}} \left(\fft{D-2}{2 (D-3)^2}\right)^{(D-3)/2},\qquad
\beta = \fft{\tilde r_+-\tilde r_-}{\tilde r_+}\alpha\,.\label{eqpars}
\ee
For neutral scalar with $q_e=0$, the equation can be solved for general $D$. For charged scalar, it can be solved only for $D=4$ and $5$.

\subsection{Neutral scalar}

Setting $q_e=0$, the equation \eqref{eqRNq} becomes
\be
x (x+1) R_{\ell m}''(x) + (2x+1) R_{\ell m}'(x)  -\tilde{\ell}(\tilde{\ell}+1) R_{\ell m}(x) = 0\,,\qquad
\tilde{\ell} = \fft{\ell}{D-3}\,.
\ee
The equation can be solved exactly in terms of hypergeometric functions. We shall not repeat the tedious process, but simply present the results. The asymptotic behavior of the solution takes the form
\bea
R_{\ell m}(x) &\sim& A x^{\tilde{\ell}} + B x^{- (\tilde{\ell}+1)}= A \left[\fft{\tilde{r}_{+}}{\tilde{r}_{+} - \tilde{r}_{-}}\right]^{\tilde{\ell}} \left(\fft{r}{r_{+}}\right)^{\ell} + B \left[\fft{\tilde{r}_{+}}{\tilde{r}_{+}-\tilde{r}_{-}}\right]^{-\tilde{\ell}-1} \left(\fft{r}{r_{+}}\right)^{-\ell-1}\,,\nn\\
&=& A' \left(\fft{r}{r_{+}}\right)^{\ell} + B' \left(\fft{r}{r_{+}}\right)^{-\ell-1}\,.
\eea
Thus the Love number is
\be
k^{(0)}_{\mathrm{RN}} = \fft{B'}{A'} = \fft{B}{A} \left(\fft{\tilde{r}_{+}-\tilde{r}_{-}}{\tilde{r}_{+}}\right)^{2\tilde{\ell}+1}
= k^{(0)}_{\mathrm{Sch}} \left( 1 - \fft{\tilde{r}_{-}}{\tilde{r}_{+}}\right)^{2 \tilde{\ell}+1}\,,
\ee
where $k^{(0)}_{\mathrm{Sch}}$ is the Love number expression for the Schwarzschild black hole \cite{Hui:2020xxx}
\be
k^{(0)}_{\mathrm{Sch}} = \fft{B}{A} = \fft{2 \tilde{\ell} + 1}{2 \pi} \fft{\Gamma (\tilde{\ell}+1)^{4}}{\Gamma (2 \tilde{\ell}+2)^{2}} \tan(\pi \tilde{\ell}) \,.
\ee
It is clear that the Love number for the RN black hole vanishes at $D=4$, but it is generally non-vanishing in other dimensions. However, it is interesting to note that in the extreme limit $\tilde{r}_{+} = \tilde{r}_{-}$, the Love number vanishes also in all dimensions. Recently there is also a simple symmetry argument which leads to the vanishing of the Love number for extreme RN black holes \cite{Kehagias:2024yzn}.

\subsection{Charged scalar in $D=5$}

For charged scalar $q_e\ne 0$, the equation \eqref{eqRNq} can be analytically solved only for $D=4$ and $D=5$. The $D=4$ case was studied earlier in the case of Kerr-Newman black hole. We now consider $D=5$, and the equation becomes
\bea
& &x (x+1) R_{\ell m}''(x) + (2x+1) R_{\ell m}'(x)  - \left(\tilde{\ell}(\tilde{\ell}+1) - \fft{(\alpha_{5} + \beta_{5} x)}{x(1+x)} \right) R_{\ell m}(x) = 0\,,\nn\\
& &\alpha_{5} = \alpha|_{D=5}\,,\qquad
\beta_{5} = \beta|_{D=5}\,,\qquad \tilde \ell=\ft12 \ell\,.
\eea
We define $ R_{\ell m}(r) = x^{s_{0}} (x+1)^{s_1} u_{\ell m}(x) $ with
\be
s_{0} = i \sqrt{\alpha_{5}}\,,\qquad
s_{1} = i \sqrt{\alpha_{5} - \beta_{5}}\,.
\ee
The function $ u $ satisfies
\be
x(1+x)\,u''_{\ell m}(x) + \left[ c x + (1+a+b-c) (1+x) \right] u'_{\ell m}(x) + a b\, u_{\ell m}(x) = 0\,,
\ee
where
\bea
a &=& -\tilde{\ell}+i\sqrt{\alpha_{5}}+i\sqrt{\alpha_{5}-\beta_{5}}\,,\qquad
b = \tilde{\ell}+1+i\sqrt{\alpha_{5}}+i\sqrt{\alpha_{5}-\beta_{5}}\,,\nn\\
c &=& 1 + 2i\sqrt{\alpha_{5} -\beta_{5}}\,.
\eea
The two linearly independent local solutions of $ u_{\ell m} $ are
\bea
u_{\ell m,1}(x) &=& {}_2F_1\left(a,b;1+a+b-c;-x\right),\nn\\
u_{\ell m,2}(x) &=& x^{c-a-b} {}_2F_1\left(c-a,c-b;1-a-b+c;-x\right).
\eea
The corresponding solutions of $ R_{\ell m}(x) $ behave at the horizon $ x=0 $ as
\bea
R_{\ell m,1}(x) \sim x^{i \sqrt{\alpha_{5}}}\,,\qquad
R_{\ell m,2}(x) \sim x^{-i \sqrt{\alpha_{5}}}\,.
\eea
As we have explained in the earlier example, we should choose choose $ R_{\ell m,1}(x) $ branch, or equivalently the $ u_{\ell m,1}(x) $. The transformation formula of hypergeometric function
\bea
&&{}_2F_1 (a,b;1+a+b-c;-x) = \fft{\Gamma(1+a+b-c) \Gamma(b-a)}{\Gamma(b) \Gamma(1+b-c)} x^{-a} {}_2F_1 (a,c-b;1+a-b;-x^{-1}) \,\nn\\
&&\qquad\qquad+ \fft{\Gamma(1+a+b-c) \Gamma(a-b)}{\Gamma(a) \Gamma(1+a-c)} x^{-b} {}_2F_1 (b,c-a;1-a+b;-x^{-1}) \,,
\eea
yields the asymptotic behavior of $ R_{1}(x) $ at the infinity:
\bea
R_{\ell m,1}(x) &\sim& \#_{1} x^{\tilde{\ell}} + \#_{2} x^{-\tilde{\ell}-1}\,,\nn\\
& \sim&  \#_{1} \left[\fft{\tilde{r}_{+}}{\tilde{r}_{+} - \tilde{r}_{-}}\right]^{\tilde{\ell}} \left(\fft{r}{r_{+}}\right)^{\ell } + \#_{2} \left[\fft{\tilde{r}_{+}}{\tilde{r}_{+}-\tilde{r}_{-}}\right]^{-\tilde{\ell}-1} \left(\fft{r}{r_{+}}\right)^{-\ell -1}\,.
\eea
We therefore obtain the Love number $ k^{(0)}_{\ell m} $:
\bea
k^{(0)}_{\ell m} &=& \fft{\Gamma(b) \Gamma(a-b) \Gamma(1+b-c)}{\Gamma(a) \Gamma(b-a) \Gamma(1+a-c)} \left(1 - \fft{\tilde{r}_{-}}{\tilde{r}_{+}}\right)^{2 \tilde{\ell} + 1}\nn\\
&=& \fft{\Gamma(-2\tilde{\ell}-1) \Gamma(\tilde{\ell}+1+i \sqrt{\alpha_{5}}-i\sqrt{\alpha_{5}-\beta_{5}})\Gamma(\tilde{\ell}+1+i\sqrt{\alpha_{5}}+i\sqrt{\alpha_{5}-\beta_{5}})}{\Gamma(2\tilde{\ell}+1) \Gamma(-\tilde{\ell}+i\sqrt{\alpha_{5}}-i\sqrt{\alpha_{5}-\beta_{5}}) \Gamma(-\tilde{\ell}+i\sqrt{\alpha_{5}}+i\sqrt{\alpha_{5}-\beta_{5}})}\,\nn\\
&\times&  \left(1 - \fft{\tilde{r}_{-}}{\tilde{r}_{+}}\right)^{2 \tilde{\ell} + 1}\,.
\eea
where $ \tilde{\ell}=\ft{\ell }{2} $, and $ \alpha_{5},\beta_{5} $ are obtained in \eqref{eqpars}. Again we see that the Love number vanishes in the extremal limit.
If $ \tilde{\ell} \in \mathbb{N} $, $ k^{(0)}_{\ell m} $ tends to pure real infinite,
\be
\fft{k^{(0)}_{\ell m}}{\Gamma(-2 \tilde{\ell}-1)} \sim - \fft{\beta_{5}}{(2\tilde{\ell})!} \prod_{n_{1}=1}^{\tilde{\ell}} \left[ n_{1}^{2} + (\sqrt{\alpha_{5}}-\sqrt{\alpha_{5}-\beta_{5}})^{2} \right] \prod_{n_{2} = 1}^{\tilde{\ell}} \left[ n_{2}^{2} + (\sqrt{\alpha_{5}}+\sqrt{\alpha_{5}-\beta_{5}})^{2} \right].
\ee
As we have seen earlier, this is a sign of the running Love number.

It turns out that in four dimensions, the charged massive scalar wave equation with generic frequency $\omega$ in the extremal RN background can be solved exactly in terms double confluent Heun's functions. This allows us to obtain formally the corresponding response coefficients. However Heun's functions can be rather complicated and we present the results in the appendix.

\section{Conclusion}

In this paper, we investigated the profile of a massless charged scalar in the background of Kerr-Newman black hole. We found that, regardless of whether the black hole is rotating, the radial equations for the static $\omega=0$ case can be expressed in the same form, namely \eqref{radial equations x}. The horizon locates at $x=0$ and spatial infinity locates at $x\rightarrow\infty$. Here $\beta$ is a parameter proportional to the scalar charge $q_e$. The $\beta$ term alters the asymptotic behavior of the equation so that the multipolar index $\ell$, which is an integer, is modified to $L$, given by \eqref{modifiedL}.

Despite having the new $\beta$ term in the scalar radial equation that is absent in the neutral case studied in literature, the scalar equation can still be solved exactly in terms of hypergeometric functions, which allows us to obtain the exact complex response coefficients: the real part gives the Love number and the imaginary part gives the dissipation number. However, the results display some important differences. We noticed that even when $\ell$ was an integer and the scalar charge $q_e$ is quantized, the hypergeometric functions would not degenerate, since $L$ is generally a real number. Hence, when charged scalar tidal force is switched on, there's no source/response ambiguity and therefore the unusual analytic continuation  $\ell\rightarrow\mathbb{R}$ is not needed.

We also observed an intriguing discontinuity of the Love number about $q_e=0$. The Love number, or the real part of the response coefficient, vanishes for the neutral scalar, but it is inversely proportional to $\beta\sim q_e$, for small but non-vanishing $q_e$. Although the divergence can be resolved by imposing charge quantization on $q_e$, the discontinuity is a curious feature of the charged response that deserves further investigation.

We also studied the near-zone of the dynamical equation of the charged scalar with non-vanishing frequency $\omega$, and found that the system had an $SL(2,\mathbb R)$ Love-like symmetry. The symmetry selects a special frequency $\omega_{\rm cr}$, which vanishes in the neutral case, so that the real part of the response coefficient vanishes identically. We refer to those dynamical states as pseudo-static.

We considered generalization of the scalar response to higher dimensions in the background of RN black holes. For the neutral scalar, the Love number is proportional to that of the Schwarzschild black hole, with a coefficient that vanishes in the extremal limit. For the charged scalar, we find that the equations can be solved exactly only in four and five dimensions. We obtained the exact complex response coefficient that would vanish in the extremal limit also. In appendix, we illustrated that the massive scalar wave equation in the extremal RN black hole background can be solved exactly in terms of double confluent Heun's functions. Our analysis provides some initial study on the charged scalar response to the charged black holes and it would also be interesting to conform our results using worldline EFT approach akin to \cite{Ivanov:2024sds}.

\section*{Acknowledgement}

We are grateful to Yue-Zhou Li for useful discussions.
L.~Ma and Y.~Pang are supported by the National Key Research and
Development Program No.~2022YFE0134300 and the National Natural Science Foundation of China (NSFC) Grant No.~12175164.
L.~Ma is also supported by Postdoctoral Fellowship Program of CPSF Grant No.~GZC20241211.
Z.-H.~Wu and H. L\"{u} are supported in part by NSFC grants No.~11935009 and No.~12375052.

\appendix

\section{Four-dimensional extremal RN black hole}

\subsection{General case with generic $(\mu_0, \omega, q_e)$}

In four dimensions, we find that the radial equation \eqref{eqRNfull} can be solved exactly in the extremal limit, even for the massive $(\mu_0 \ne 0)$ and dynamic $(\omega \ne 0)$ charged $(q_e\ne 0)$ scalar. Defining a new radial variable $z=\fft{r-r_h}{r_h} $, where $r_h$ is the horizon of the extremal black hole, the equation \eqref{eqRNfull} can be transformed into
\be
R''(z) + \fft{2}{z} R' (z) + \left(E_0 + \frac{E_1}{z} + \frac{E_2}{z^2} + \frac{E_3}{z^3} + \frac{E_4}{z^4}\right) R(z) = 0\,,\label{eqRNz}
\ee
where the coefficients are
\bea
E_0 &=& r_h^2(\omega^2 - \mu_0^2)\,,\quad E_1 = 2 r_h^2 (2 \omega^2 - q_e \omega - \mu_0^2)\,,\quad
E_2 = r_h^2(6 \omega^2 - 6 q_e \omega + q_e^2 - \mu_0^2) -\ell(\ell+1) \,,\nn\\
E_3 &=& 2 r_h^2 (2 \omega^2 - 3 q_e \omega - q_e^2)\,,\qquad
E_4 = r_h^2 (\omega - q_e)^2\,.
\eea
We further re-scale $ R(z) = z^{-\frac{1}{2}} u(\tilde{z}) $ and define $ \tilde{z} = (\frac{E_0}{E_4})^\fft14  z$. The radial equation becomes the double confluent Heun equation:
\be
u''(\tilde{z}) + \tilde z^{-1} u'(\tilde{z}) + \tilde z^{-2} (-\kappa^2 \tilde{z}^2 + b \tilde{z} - c^2 + d \tilde{z}^{-1} - \kappa^2 \tilde{z}^{-2}) u(\tilde{z}) = 0\,.\label{dcheN2}
\ee
The parameters are shown below
\bea
\kappa &=& - i E_0^\fft14 E_4^\fft14 = - i r_h \sqrt{\omega - q_e}(\omega^2 - \mu_0^2)^\fft14\,,\nn\\\
b &=& E_1 (\fft{E_4}{E_0})^\fft14 = 2 r_h^2\fft{\sqrt{\omega- q_e } (2 \omega^2 - q_e \omega -\mu_0^2)}{(\omega^2 - \mu_0^2)^\fft14}\,,\nn\\
d &=& E_3 (\fft{E_0}{E_4})^\fft14 = 2 r_h^2 \fft{(\omega^2 - \mu_0^2)^\fft14 (2 \omega^2 - 3 q_e \omega + q_e^2)}{\sqrt{\omega - q_e}}\,,\nn\\
c &=& \fft12 \sqrt{1-4 E_2} = \fft12 \sqrt{1 + 4\ell(\ell+1) - 4 r_h^2 (6 \omega^2 - 6 q_e \omega + q_e^2 - \mu_0^2)}\,.
\eea
The Heun functions are much more complicated than the hypergeometric functions that appeared earlier. We make extensive use of Ref.~\cite{buhring_double_1994} on the Heun functions. From now on, we shall simply drop the tilde symbol on $\tilde z$ for simplicity. The two local solutions on the horizon $z=0$ and asymptotic infinity $z \rightarrow \infty$ behave respectively as
\bea
u_{01}(z) &=& \exp(\frac{\kappa}{z}) z^{\ft12 +  \frac{d}{2\kappa}} \sum_{n=0}^{\infty} a_n(d,b,\kappa) n! (\fft{z}{2 \kappa})^n\,,\nn\\
u_{02}(z) &=& \exp(-\frac{\kappa}{z}) z^{\ft12 - \frac{d}{2\kappa}} \sum_{n=0}^{\infty} a_n(d,b,-\kappa) n! (-\fft{z}{2 \kappa})^n\,,\nn\\
u_{\infty 1}(z) &=& \exp(\kappa z) z^{-\ft12 -\frac{b}{2\kappa}} \sum_{n=0}^{\infty} a_n(b,d,\kappa) n! (2 \kappa z)^{-n}\,,\nn\\
u_{\infty 2}(z) &=& \exp(\kappa z) z^{-\ft12 +\frac{b}{2\kappa}} \sum_{n=0}^{\infty} a_n(b,d,-\kappa) n! (-2 \kappa z)^{-n}\,,
\eea
where $a_n = a_n(b,d,\kappa)$ satisfies a four-term recurrence relation
\be
a_n = \fft{1}{n^2} \Big[ (-c-\fft12+\frac{b}{2\kappa}+n)(c-\fft12+\frac{b}{2\kappa}+n) a_{n-1}+\fft{2 \kappa d}{n-1} a_{n-2} - \fft{4 \kappa^4}{(n-2)(n-1)}a_{n-3} \Big].\label{an}
\ee
Note that we use $z$ instead of $\tilde{z}$ in the function $u$ for simplicity. Furthermore, we have
\bea
z\rightarrow 0:&&\qquad u_{01} \approx \exp(- i \frac{\#}{z})\,,\qquad
u_{02} \approx \exp( i \frac{\#}{z})\,,\nn\\
z\rightarrow \infty:&&\qquad u_{\infty 1} \approx \exp(- i \# z)\,,\qquad
u_{\infty 2} \approx \exp( i \# z)\,,
\eea
where ``$\#$'' denotes constant coefficients whose explicit expressions are unimportant for the discussion. It is thus clear that $u_{01}$ is the ingoing mode on the horizon and $u_{\infty 1}$, $u_{\infty 2}$ are transmission and reflecting waves respectively at the asymptotic infinity.

We choose the ingoing mode on the horizon and the natural generalization of the Love number to non-dynamic case is the ratio of reflecting wave to transmission wave, which is actually the transmission coefficient. The local solutions at on the horizon and asymptotic infinity are connected as $u_{01}(z) = Q_{11} u_{\infty 1}(z) + Q_{12} u_{\infty 2}(z) $ \cite{buhring_double_1994}, so that the generalized Love number is
\be
k = \fft{Q_{12}}{Q_{11}}\,,\label{kford4extreme}
\ee
where $ Q_{11}$, $Q_{12} $ were obtained in \cite{buhring_double_1994}, given by
\bea
Q_{11} &=& \kappa^{\frac{d-b}{2\kappa}} \exp[i \pi (\ft12 - \ft{d}{2\kappa})] \fft{e(d,b,-\kappa)}{\cos(\pi \nu)} \fft{1}{2 \sin(\pi \nu)}\,\nn\\
& &\times \bigg\{ - \sqrt{\fft{e(b,d,\kappa)}{\cos[ \pi (\nu + \frac{b}{2\kappa})]}} \sqrt{\fft{e(d,b,\kappa)}{\cos[\pi (-\nu +\frac{d}{2\kappa})]}} r_\nu \exp(i \pi \nu)\,\nn\\
& & + \sqrt{\fft{e(b,d,\kappa)}{\cos[\pi (-\nu + \frac{b}{2\kappa})]}} \sqrt{\fft{e(d,b,\kappa)}{\cos[\pi (\nu +\frac{d}{2\kappa})]}} r_{-\nu} \exp(-i \pi \nu) \bigg\}\,,\nn\\
Q_{12} &=& \kappa^{ \frac{b+d}{2\kappa}}\exp[ - i \pi \ft{b+d}{2 \kappa} ] \fft{e(d,b,-\kappa)}{\cos(\pi \nu)} \fft{1}{2 \sin(\pi \nu)}\nn\\
& & \times \bigg\{ \sqrt{\fft{e(b,d,-\kappa)}{\cos[ \pi (\nu -\frac{b}{2\kappa}) ]}} \sqrt{\fft{e(d,b,\kappa)}{\cos [ \pi (-\nu +\frac{d}{2\kappa}) ]}} r_\nu \exp( 2 i \pi \nu )\,\nn\\
& & - \sqrt{\fft{e(b,d,-\kappa)}{\cos[ \pi (-\nu -\frac{b}{2\kappa}) ]}} \sqrt{\fft{e(d,b,\kappa)}{\cos[ \pi (\nu +\frac{d}{2\kappa}) ]}}r_{-\nu} \exp( -2 i \pi \nu )\bigg\}\,,\label{connectCoe}
\eea
where the parameters $(e, \nu, r_{\pm\nu})$ need to be further clarified.  The quantity $e(b,d,\kappa)$ is the solution of a system of linear equation, see (2.19) and (2.34) in \cite{buhring_double_1994} for detailed expression, and we simplify the notations as $ e(\kappa) = e(b,d,\kappa)$, $\tilde{e}(\kappa) = e(d,b,\kappa) $. There is  a limit formula for $e$ as
\be
e(-\kappa) = \pi \lim_{n \rightarrow \infty} \fft{(n!)^2}{\Gamma (c + \fft12 +\frac{b}{2\kappa} +n ) \Gamma(-c + \fft12 +\frac{b}{2\kappa} + n )}2^{\frac{b}{\kappa}} a_n(b,d,\kappa)\,,\label{ek}
\ee
and $\nu$ is the characteristic exponent determined by
\be
\cos^2( \pi \nu ) = \sin^2 \left( \fft{\pi b}{2 \kappa} \right) + e(-\kappa)\, e(\kappa)\,.\label{mu}
\ee
The quantities $ r_{\pm \nu}$ are defined by
\be
r_{\mu} = \fft{\Delta_{-\mu}(d,b)}{\Delta_{\mu}(b,d)}\kappa^{-2\mu}\,,\qquad
\mu \in \{ -\nu,\nu \}\,,\label{rnu}
\ee
where $\Delta_{\mu}(b,d)$ and $\Delta_{-\mu}(d,b) $ are given by
\bea
\Delta_{\mu} (b,d) &=& \sqrt{\fft{e(\kappa)}{\cos[ \pi (\mu + \frac{b}{2\kappa}) ]}} \sum_{l=0}^{\infty} \fft{l!}{\Gamma(\mu +\ft32 +\frac{b}{2\kappa} + l)} 2^{-l} a_{l}(b,d,\kappa)\cr
& & + \sqrt{\fft{e(-\kappa)}{\cos[ \pi (\mu - \frac{b}{2\kappa}) ]}} \sum_{l=0}^{\infty} \fft{l!}{\Gamma(\mu +\ft32 -\frac{b}{2\kappa} + l)} 2^{-l} a_l(b,d,-\kappa)\,,\nn\\
\Delta_{-\mu} (d,b) &=& \sqrt{\fft{\tilde{e}(\kappa)}{\cos[ \pi (-\mu +\frac{d}{2\kappa}) ]}} \sum_{l=0}^{\infty} \fft{l!}{\Gamma(-\mu +\ft32 +\frac{d}{2\kappa} + l )} 2^{-l} a_l(d,b,\kappa)\cr
& & + \sqrt{\fft{\tilde{e}(-\kappa)}{\cos[ \pi (-\mu -\frac{d}{2\kappa}) ]}} \sum_{l=0}^{\infty} \fft{l!}{\Gamma(-\mu + \ft32 - \frac{d}{2\kappa} + l )}2^{-l}a_l (d,b,-\kappa)\,.\label{delta}
\eea
There should be no confusing of this $\mu \in \{-\nu, \nu\}$ to the mass parameter of the black hole and scalar mass $\mu_0$.

\subsection{Special case: $\mu_0=0$, $\omega= \ft12 q_e$ }

In the previous subsection, we obtain formally the expression of the generalized Love number \eqref{kford4extreme}. However, the expressions of various parameters are so complex that we cannot have a clear cut answer.
The expressions can be somewhat simplified when $b=d=0$, as discussed in section 8 of \cite{buhring_double_1994}. The $b=d=0$ case can be achieved by setting $\mu_0=0$ and $\omega = \ft12 q_e$. In this case, it follows from \eqref{an} that we now have a simpler recursion
\be
a_n = \fft{1}{n^2} \left[ (-c-\ft12+n)(c-\ft12+n)a_{n-1} - \fft{4 \kappa^4}{(n-2)(n-1)}a_{n-3} \right],
\ee
where $a_n\equiv a_n(0,0,\kappa)=a_n(0,0,-\kappa)$. The exponent $\nu$ now satisfies
\be
\cos\pi\nu = e \equiv e(0,0,\kappa)=e(0,0,-\kappa)\,.\label{nue}
\ee
The parameters $r_\mu$'s ($\mu=\pm \nu$) satisfy
\be
r_{\mu} = \fft{\Delta_{-\mu}}{\Delta_{\mu}}\kappa^{-2 \mu}\,,\qquad
r_{\nu} r_{-\nu} =1\,,\qquad
\Delta_\mu = 2 \sum_{n=0}^{\infty} \fft{n!}{\Gamma( \mu + \ft32 +n )} 2^{-n} a_n\,.
\ee
The generalized Love number \eqref{kford4extreme} now becomes much simpler, given by
\be
k = i \fft{e^{2 i \pi \nu} r_\nu - e^{-2 i \pi \nu} r_{-\nu}}{e^{i \pi \nu} r_\nu - e^{-i \pi \nu} r_{-\nu}}\,.
\ee
This expression no longer explicitly depends on $e$, but it still depends on $e$ through via \eqref{nue}.

\end{document}